\pdfoutput=1
\documentclass[aps,pre,reprint,groupedaddress]{revtex4-1}

\usepackage[final]{graphicx}
\usepackage{amsmath}
\usepackage{verbatim}
\usepackage[colorlinks,
linkcolor=blue,
citecolor=blue,
urlcolor=blue]{hyperref}

\begin{document}

\preprint{}

\title{Fluid nonlinear frequency shift of nonlinear ion acoustic waves in multi-ion species plasmas in small wave number region}

\author{Q. S. Feng} 
\affiliation{HEDPS, Center for
	Applied Physics and Technology, Peking University, Beijing 100871, China}

\author{C. Z. Xiao}
\affiliation{HEDPS, Center for
	Applied Physics and Technology, Peking University, Beijing 100871, China}

\author{Q. Wang}
\affiliation{HEDPS, Center for
	Applied Physics and Technology, Peking University, Beijing 100871, China}

\author{C. Y. Zheng} \email{zheng\_chunyang@iapcm.ac.cn}

\affiliation{HEDPS, Center for
	Applied Physics and Technology, Peking University, Beijing 100871, China}
\affiliation{Institute of Applied Physics and Computational
	Mathematics, Beijing, 100094, China}
\affiliation{Collaborative Innovation Center of IFSA (CICIFSA) , Shanghai Jiao Tong University, Shanghai, 200240, China}

\author{Z. J. Liu} 
\affiliation{HEDPS, Center for
	Applied Physics and Technology, Peking University, Beijing 100871, China}
\affiliation{Institute of Applied Physics and Computational
	Mathematics, Beijing, 100094, China}

\author{L. H. Cao} 
\affiliation{HEDPS, Center for
	Applied Physics and Technology, Peking University, Beijing 100871, China}
\affiliation{Institute of Applied Physics and Computational
	Mathematics, Beijing, 100094, China}
\affiliation{Collaborative Innovation Center of IFSA (CICIFSA) , Shanghai Jiao Tong University, Shanghai, 200240, China}
	
\author{X. T. He}
\affiliation{HEDPS, Center for
	Applied Physics and Technology, Peking University, Beijing 100871, China}
\affiliation{Institute of Applied Physics and Computational
	Mathematics, Beijing, 100094, China}
\affiliation{Collaborative Innovation Center of IFSA (CICIFSA) , Shanghai Jiao Tong University, Shanghai, 200240, China}

\date{\today}

\begin{abstract}
The properties of the nonlinear frequency shift (NFS) especially the fluid NFS from the harmonic generation of the ion-acoustic wave (IAW) in multi-ion species plasmas have been researched by Vlasov simulation.
The pictures of the nonlinear frequency shift from harmonic generation and particles trapping are shown to explain the mechanism of NFS qualitatively. The theoretical model of the fluid NFS from harmonic generation in multi-ion species plasmas is given and the results of Vlasov simulation are consistent to the theoretical result of multi-ion species plasmas. When the wave number $k\lambda_{De}$ is small, such as $k\lambda_{De}=0.1$, the fluid NFS dominates in the total NFS and will reach as large as nearly $15\%$ when the wave amplitude $|e\phi/T_e|\sim0.1$, which indicates that in the condition of small $k\lambda_{De}$, the fluid NFS dominates in the saturation of stimulated Brillouin scattering especially when the nonlinear IAW amplitude is large.

\end{abstract}


\maketitle

\section{Introduction}
Stimulated Brillouin scattering (SBS), or resonant decay of a light wave into a scattered light wave and an ion acoustic wave (IAW),\cite{Kruer} plays an important role in the successful ignition goal of inertial confinement fusion (ICF).\cite{(1), (9)_37}
 Many methods have been taken to reduce the SBS scattering level,\cite{Liu_8,(9)_6,(9)_49,(9)_7, (9)_46} including frequency detuning due to particle-trapping,\cite{(9)_6_2,(9)_6_2_1,(9)_6_4,(9)_6_5,(9)_6_5_1,(9)_6_5_2} increasing linear Landau damping induced by kinetic ion heating,\cite{(9)_6_9,(9)_6_10} nonlinear damping due to wave-breaking and trapping\cite{(9)_6_11,(9)_6_12,(9)_6_12_1,(9)_6_12_2} or coupling with higher harmonics\cite{(9)_6_13,(9)_6_14}.

The nonlinear effect on the frequency and Landau damping rate of nonlinear ion acoustic waves have received renewed interest for their potential role in determining the saturation in stimulated Brillouin scattering of laser drivers in inertial confinement fusion (ICF).\cite{Cohen_1997POP}-\cite{Cohen_2005POP} Recently, Albright et al. researched ion-trapping-induced and electron-trapping-induced IAW bowing and breakup, they found SBS saturation was dominated by IAW bowing and breakup due to nonlinear frequency shift from particles trapping.\cite{Albright_2016POP} However, this paper will show if the wave number $k\lambda_{De}$ is small, the fluid nonlinear frequency shift from harmonic generation will dominate and should be considered especially when the wave amplitude is large.

The nonlinear behavior of an ion acoustic wave is a subject of fundamental interest to plasma physics. The total nonlinear frequency shift (NFS) comes from the harmonic generation and particles trapping. In 1997, the harmonic nonlinearities in IAW were researched by Cohen et al. through the single specie cold ion fluid equations.\cite{Cohen_1997POP} Then, in 2007, the fluid NFS due to harmonic nonlinearities was reinvestigated by Winjum et al. although the research was about the electron plasma wave (EPW).\cite{Winjum_2007POP} The fluid frequency shift is positive and proportional to the square of the wave amplitude. In 2013, Berger et al. derived the fluid NFS of IAW in single-ion specie plasmas from the isothermal cold ion fluid equations.\cite{(9)} This paper will show the fluid NFS of IAW in multi-ion species plasmas.

In the early 1970s, the kinetic nonlinear frequency shift (KNFS) due to particles trapping was calculated by Dewar\cite{Dewar} and Morales and O'Neil\cite{Morales}. In their work, they assumed that the wave did not contain harmonics and these derivations were for the case of an EPW, and were applied to IAW by Berger et al.\cite{(9)}. The KNFS is proportional to the square root of the wave amplitude. In 2013, Chapman et al. researched the kinetic nonlinear frequency shift in multi-ion species plasmas.\cite{T. Chapman_PRL} In their research, in the condition of the wave number $k\lambda_{De}=1/3$, the KNFS dominated in the scope of wave amplitude researched. However, the fluid nonlinear frequency shift from harmonic generation made use of the result of single-ion specie plasmas that was derived by Berger\cite{(9)}. Therefore, it was clear that calculations of the KNFS matched Vlasov results well for low wave amplitude, but underestimated the total NFS at higher amplitudes where harmonic generation was expected to contribute a further positive frequency shift. For the system in Chapman's research is multi-ion species system, the fluid NFS model should make use of the multi-ion species fluid NFS model. This paper will give a multi-ion species fluid NFS model to improve the single-ion specie fluid NFS model. And the multi-ion species fluid NFS model have a further positive frequency shift than the single-ion specie fluid NFS model and will fit the Vlasov results better.

In this paper, when the wave number is small such as $k\lambda_{De}=0.1$, the fluid NFS from harmonic generation is important especially when the wave amplitude is large. And the multi-ion species fluid NFS model is given to calculate the fluid NFS. The Vlasov simulation results are verified to be consistent to the theoretical model of the total NFS including fluid NFS and kinetic NFS in multi-ion species plasmas. When $k\lambda_{De}=0.1$, the total NFS can reach a value nearly $15\%$ when the nonlinear IAW amplitude $|e\phi/T_e|$ reaches about 0.1, which is much larger than the total NFS in the condition of $k\lambda_{De}=1/3$\cite{T. Chapman_PRL}. This indicates that when $k\lambda_{De}$ is small, the fluid NFS will reach a large level especially when the wave amplitude is large, thus dominating in the total NFS so that the fluid NFS dominants in the saturation of SBS by IAW bowing and breakup\cite{Albright_2016POP}.

\section{\label{Sec:Theory model}Theoretical model}

\subsection{\label{Subsec:A. Fluid theory}Fluid theory: nonlinear frequency shift from harmonic generation in multi-ion species plasmas}

Taking the charge and the number fraction of the ion specie $\alpha$ to be $Z_\alpha$ and $f_\alpha$, the electron density, $n_e$, and the total ion density, $n_i$, are related through the average charge, $\bar{Z}$, i.e., $n_e=n_i\bar{Z}$. The ion number densities are given by 
\begin{equation}
\begin{aligned}
n_{\alpha}=f_\alpha n_i=f_\alpha n_e/\bar{Z},\\
\bar{Z}=\sum\limits_{\alpha}f_\alpha Z_\alpha.
\end{aligned}
\end{equation}

Here, multi ion species and isothermal Boltzmann electrons are considered in this cold ion fluid model, the nonlinear frequency shift from harmonic generation is derived from the multi species isothermal cold ions fluid equations:

\begin{equation}
\label{Eq:Fluid}
\begin{aligned}
&\frac{\partial n_\alpha}{\partial t}+\frac{\partial n_\alpha v_\alpha}{\partial x}=0,\\
&\frac{\partial v_\alpha}{\partial t}+v_\alpha \frac{\partial v_\alpha}{\partial x}=-C_\alpha^2\frac{\partial \tilde{\phi}}{\partial x},\\
&-\frac{T_e}{e}\frac{\partial^2\tilde{\phi}}{\partial x^2}+4\pi en_{e0}\text{exp}(\tilde{\phi})=4\pi \sum\limits_{\alpha}q_\alpha n_\alpha,
\end{aligned}
\end{equation}
where $C_\alpha=\sqrt{Z_\alpha T_e/m_\alpha}$ is the sound velocity of ion $\alpha$, $\tilde{\phi}=e\phi/T_e$ is the normalized electrostatic potential. Initial charge satisfies charge conservation, i.e., $\sum\limits_{\alpha}q_\alpha n_{\alpha0}=en_{e0}$. Following Berger et al.,\cite{(9)} one expands the variables in a Fourier series
\begin{equation}
\begin{aligned}
(\tilde{\phi}, n_\alpha, v_\alpha)=&(\tilde{\phi}_0, n_{\alpha 0}, v_{\alpha 0})+\\
&\frac{1}{2}\sum\limits_{l\neq 0}(\tilde{\phi}_l, n_{\alpha l}, v_{\alpha l})\text{exp}[il(kx-\omega t)],
\end{aligned}
\end{equation}
with the condition $(\tilde{\phi}_{-l}, n_{\alpha,-l}, v_{\alpha,-l})=(\tilde{\phi}_l, n_{\alpha l}, v_{\alpha l})^*$ and keeping terms for $l=0, \pm1, \pm2$ up to 2nd order with $\text{exp}  \tilde{\phi}\simeq1+\tilde{\phi}+\frac{1}{2}\tilde{\phi}^2$.
Retaining the same exponents in the Fourier series, for $v_0=0, n_{\alpha 0}=f_\alpha n_{i0}=f_\alpha n_{e0}/\bar{Z}=\text{constant}$, one finds from Eq. (\ref{Eq:Fluid}) for $l=0$
\begin{equation}
en_{e0}(1+\tilde{\phi}_0+\frac{1}{2}\tilde{\phi}_0^2+\frac{1}{4}|\tilde{\phi}_1|^2+\frac{1}{4}|\tilde{\phi}_2|^2)=\sum\limits_{\alpha}q_\alpha n_{\alpha 0},
\end{equation}
by conservation of charge, $en_{e0}=\sum\limits_{\alpha}q_\alpha n_{\alpha 0}$, it thus obtains that
\begin{equation}
\tilde{\phi}_0+\frac{1}{2}\tilde{\phi}_0^2=-\frac{1}{4}|\tilde{\phi}_1|^2-\frac{1}{4}|\tilde{\phi}_2|^2.
\end{equation}

For $l=1$, the equations are
\begin{equation}
\begin{aligned}
&-i\omega n_{\alpha 1}+ikn_{\alpha 0}v_{\alpha 1}=-i\frac{1}{2}kn_{\alpha,-1}v_{\alpha 2}-i\frac{1}{2}kn_{\alpha 2}v_{\alpha,-1},\\
&-i\omega v_{\alpha 1}+ikC_\alpha^2\tilde{\phi}_1=-i\frac{1}{2}kv_{\alpha,-1}v_{\alpha 2},\\
&[\frac{T_e}{e}k^2+4\pi en_{e0}(1+\tilde{\phi}_0)]\tilde{\phi}_1  \\ &-4\pi\sum\limits_{\alpha}q_{\alpha}n_{\alpha 1}=-\frac{1}{2}4\pi en_{e0}\tilde{\phi}_2\tilde{\phi}_{-1}.
\end{aligned}
\end{equation}

For $l=2$, the corresponding equations are
\begin{equation}
\begin{aligned}
&-2i\omega n_{\alpha 2}+i2kn_{\alpha 0}v_{\alpha2}+ikn_{\alpha 1}v_{\alpha 1}=0,\\
&-2i\omega v_{\alpha 2}=-i2k[C_{\alpha}^2\tilde{\phi}_2+\frac{1}{4}v_{\alpha 1}^2],\\
&[4\frac{T_e}{e}k^2+4\pi en_{e0}(1+\tilde{\phi}_0)]\tilde{\phi}_2+\frac{1}{4}4\pi en_{e0}\tilde{\phi}_1^2=4\pi\sum\limits_{\alpha}q_\alpha n_{\alpha 2}.
\end{aligned}
\end{equation}

Defining $C_{2s}^2=\sum\limits_{\alpha}\frac{Z_\alpha n_{\alpha 0}}{n_{e0}}C_\alpha^2$,   $C_{4s}^4=\sum\limits_{\alpha}\frac{Z_\alpha n_{\alpha 0}}{n_{e0}}C_\alpha^4$ and $C_{sl}^2=l^2k^2C_{2s}^2/(1+l^2k^2\lambda_{De}^2)$, by keeping terms only to 2nd order in $\tilde{\phi}_1$, then one obtains the relation between $\tilde{\phi}_2$ and $\tilde{\phi}_1$:
\begin{equation}
\label{Eq:phi_2-phi_1}
\tilde{\phi}_2=A_{2\phi}\tilde{\phi}_1^2,
\end{equation}
where 
\begin{equation}
\label{Eq:A_2}
A_{2\phi}=\frac{1}{4}\frac{C_{s2}^2}{4\omega^2-C_{s2}^2}[3\frac{k^2C_{4s}^4}{\omega^2C_{2s}^2}-\frac{\omega^2}{k^2C_{2s}^2}].
\end{equation}

Using $\tilde{\phi}_0\simeq-|\tilde{\phi}_1|^2/4$ and defining $C_{6s}^6=\sum\limits_{\alpha}Z_\alpha n_{\alpha 0}C_\alpha^6/n_{e0}$ correspondingly, retaining harmonic corrections to the linear dispersion relation again to the second order in $\tilde{\phi}_1$ and keeping the lowest nonlinear term, the nonlinear equation for the amplitude of $\tilde{\phi}_1$ is obtained as follows:
\begin{equation}
(\omega^2-C_{s1}^2)\tilde{\phi}_1=(A_{2\phi}C_{A_{2\phi}}+C_2)|\tilde{\phi}_1|^2\tilde{\phi}_1,
\end{equation}
where
\begin{equation}
\begin{aligned}
\label{Eq:C_A}
&C_{A_{2\phi}}=\frac{C_{s1}^2}{2}[-\frac{\omega^2}{k^2C_{2s}^2}+3\frac{k^2C_{4s}^4}{\omega^2C_{2s}^2}],\\
&C_2=\frac{C_{s1}^2}{8}[2\frac{\omega^2}{k^2C_{2s}^2}+5\frac{k^4C_{6s}^6}{\omega^4C_{2s}^2}].
\end{aligned}
\end{equation}

Taking the effective fundamental IAW frequency after accounting for harmonic effects $\omega_{harm}=\omega_0+\delta \omega_{harm}$ and the linear fundamental frequency of IAW $\omega_0\simeq C_{s1}={kC_{2s}}/\sqrt{1+k^2\lambda_{De}^2}=kC_\alpha\sqrt{\sum\limits_{\alpha}\frac{Z_\alpha n_{\alpha 0}}{n_{e0}}}/\sqrt{1+k^2\lambda_{De}^2}$, then the frequency shift of the first harmonic due to the inclusion of the second harmonic terms in multi species ions plasmas is given by 
\begin{equation}
\label{Eq:fluid_frequencyShift}
\frac{\delta\omega_{harm}}{\omega_0}=\frac{1}{2}\frac{\Delta}{\omega^2}|\tilde{\phi}_1|^2=L|\tilde{\phi_1}|^2,
\end{equation}
where $L=\frac{1}{2}\frac{\Delta}{\omega^2}$ is the fluid NFS coefficient and $\Delta$ is defined by
\begin{equation}
\label{Eq:Delta}
\Delta\equiv A_{2\phi}C_{A_{2\phi}}+C_2.
\end{equation}
By combining Eqs. (\ref{Eq:C_A}), (\ref{Eq:fluid_frequencyShift}), (\ref{Eq:Delta}), 
then the final result of $\delta\omega_{harm}/\omega$ is related to the wave number $k\lambda_{De}$ and the second order in $\tilde{\phi}_1$ 
\begin{equation}
\begin{aligned}
&\frac{\delta\omega_{harm}}{\omega_0}=L|\tilde{\phi_1}|^2=\\
&[\frac{-2A_{2\phi}+1}{8(1+k^2\lambda_{De}^2)}+\frac{(1+k^2\lambda_{De}^2)^2}{(C_{2s}^2)^2}(\frac{6A_{2\phi} C_{4s}^4}{1+k^2\lambda_{De}^2}+5\frac{C_{6s}^6}{C_{2s}^2})]|\tilde{\phi_1|^2},
\end{aligned}
\end{equation} \\
where the 2nd harmonic coefficient $A_{2\phi}$ is defined in Eq. (\ref{Eq:A_2}). 

\begin{figure}[!tp]
	\includegraphics[width=1.0\columnwidth]{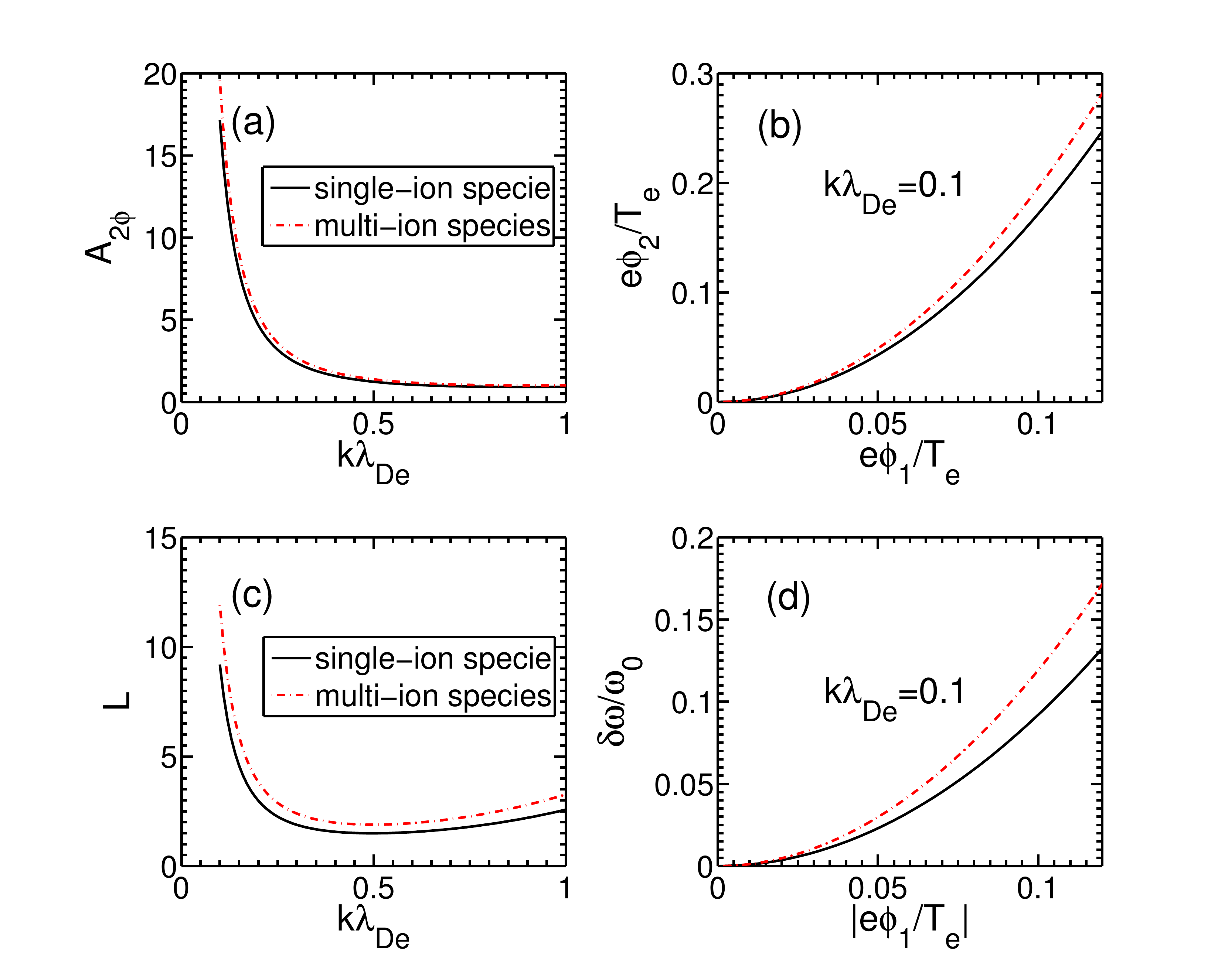}
	
	\caption{\label{Fig:single-multi}(Color online) (a) The coefficient of the 2nd harmonic $A_{2\phi}$ and (c) the fluid NFS coefficient $L$ as a function of the wave number $k\lambda_{De}$. In the condition of $k\lambda_{De}=0.1$, (b) the 2nd harmonic amplitude $e\phi_2/T_e$ and (d) the fluid NFS $\delta\omega_{harm}/\omega_0$ as a function of $|e\phi_1/T_e|$. The black solid line presents for single-ion specie plasmas and the red dot dash line for CH (1:1) plasmas.}
\end{figure}

As shown in Fig. \ref{Fig:single-multi}(a), when $k\lambda_{De}$ is small, the 2nd harmonic coefficient $A_{2\phi}$ is large and the contribution of $\tilde{\phi}_2$ is strong for the reason that the second harmonic becomes more resonant as $4\omega^2-C_{s2}^2=4C_{s1}^2-C_{s2}^2$ decreases strongly with $k\lambda_{De}$ decreasing. Fig. \ref{Fig:single-multi}(c) shows that with $k\lambda_{De}$ increasing, the fluid NFS coefficient first decreases quickly and then slowly increases, the turning point is $k\lambda_{De}\simeq0.5$ and the SBS relevant wave number is usually $k\lambda_{De}<0.5$. To research the fluid NFS in multi-ion species plasmas, CH (1:1) plasmas are taken as an typical multi-ion species plasmas. We can see that the 2nd harmonic coefficient $A_{2\phi}$ of CH plasmas is slightly larger than that of the single-ion specie plasmas, as a result, the fluid NFS coefficient $L$ from the harmonic generation in CH plasmas is slightly higher than that of the single-ion specie plasmas as shown in Figs. \ref{Fig:single-multi}(a) and \ref{Fig:single-multi}(c). Especially, when $k\lambda_{De}$ is small, such as $k\lambda_{De}=0.1$, harmonic effect is obvious and the deviation of the harmonic effect between the single-ion specie plasmas and the multi-ion species plasmas is also obvious as shown in Figs. \ref{Fig:single-multi}(b) and \ref{Fig:single-multi}(d). When the number fraction $f_H=0$ or $f_C=0$, the results from the multi-ion species plasmas including $A_{2\phi}$ and $L$ are the same with that of the single-ion specie plasmas. When the number fraction varies, the results from the multi-ion species plasmas including $A_{2\phi}$ and $L$ varies correspondingly.

\subsection{\label{Subsec:B. Kinetic theory}Kinetic theory: nonlinear frequency shift from particles trapping in multi-ion species plasmas}

The linear kinetic theory of ion-acoustic wave (IAW) in homogeneous, non-magnetized plasmas consisting of multi-species ions is considered. Assume that the same temperature of all ions equals to $T_i$ and the electron temperature equals to $T_e$. The linear dispersion relation of the IAW in multi-ion species  plasmas is given by
\begin{equation}
\epsilon_L(\omega,k)=1+\sum_j \frac{1}{(k\lambda_{Dj})^2}(1+\xi_jZ(\xi_j))=0,
\end{equation}
where $\xi_j=\omega/(\sqrt{2}kv_{tj})$, and $Z(\xi_j)$ is the dispersion function, j presents for electron, H ion or C ion.  $v_{tj}=\sqrt{T_j/m_j}$ is the thermal velocity of particle $j$. $\lambda_{Dj}=\sqrt{T_j/4\pi n_jZ_j^2e^2}$ is the Debye length. $T_j, m_j, n_j, Z_j$ are the temperature, mass, density and charge number of specie j.

The derivation of the kinetic nonlinear frequency shift (KNFS) of electrostatic waves resulting from particles trapping is originally presented by Dewar\cite{Dewar} and Morales and O'Neil\cite{Morales}, and is applied to the case of IAWs in single ion specie plasmas (such as H or He ion) by Berger\cite{(9)}.  The following kinetic nonlinear frequency shift (KNFS) in multi-ion species plasmas is given by Berger:
\begin{equation}
\label{Eq:KNFS}
\delta\omega_{NL}^{kin}\simeq-[\frac{\partial\epsilon_L(\omega_L)}{\partial{\omega}}]^{-1}\sum_j \alpha_j\frac{\omega_{pj}^2}{k^2}\Delta v_{tr,j}\frac{d^2(f_0/N)}{dv^2}\rvert_{v_\phi},
\end{equation}
 where the initial distribution $f_0$ is assumed to be Maxwellian distribution, then
$$
 \frac{d^2(f_0/N)}{dv^2}=\frac{1}{\sqrt{2\pi}}\frac{1}{v_{th}^3}[(v/v_{th})^2-1]e^{-\frac{1}{2}(\frac{v}{v_{th}})^2},
 $$
and
$ \Delta v_{tr,j}=2\sqrt{\frac{|q_j|\phi_0}{m_j}}
$,
 $\alpha$ is a constant which is dependent on how the wave was excited, $\alpha_j=\alpha_{sud}=0.823$ for sudden excitation, $\alpha_j=\alpha_{ad}=0.544$ for adiabatic excitation, j presents for H ion or C ion, however, the excitation condition of electrons is adiabatic i.e., $\alpha_e=\alpha_{ad}=0.544$.\cite{(9)} In this paper, adiabatic excitation is taken for it is more relevant to the conditions here and stimulated Brillouin scattering processes.

\section{\label{Sec:Vlasov simulation}Vlasov simulation}

One dimension in space and velocity (1D1V) Vlasov-Poisson code\cite{Liu_2009POP,Liu_2009POP_1} is taken to excite the nonlinear IAW and research the properties of the nonlinear frequency shift of the nonlinear IAW in CH plasmas. The split method\cite{7_5-1976JCP,7-2004CPC} is taken to solve Vlasov equation, we split the time-stepping operator into free-streaming in x and motion in $v_x$, then we can get the advection equations. A third order Van Leer scheme (VL3)\cite{VL3, 10-2006POP} is taken to solve the advection equations.  A neutral, fully ionized, non-magnetized CH plasmas (1:1 mixed) with the same temperature of all ion species ($T_H=T_C=T_i$) is considered. To solve the particles behavior, the phase space domain is $[0, L_x]\times[-v_{max}, v_{max}]$, where $L_x=2\pi/k$ is the spatial scale discretized with $N_x=128$ grid points in the spatial domain and $v_{max}=8v_{tj}$ (j presents for electrons, H ions or C ions) is the velocity scale with $N_v=256$ grid points in the velocity domain. The periodic boundary condition is taken in the spatial domain and the time step is $dt=0.1\omega_{pe}^{-1}$. To generate a driven IAW, one considers the form of the external driving electric field (driver) as
\begin{equation}
\tilde{E}_d(x, t)=\tilde{E}_d(t)\text{sin}(kx-\omega_d t),
\end{equation}
where tilde presents for normalization $\tilde{E}_d=eE_d\lambda_{De}/T_e$. $\omega_d$ and k are the frequency and the wave number of the driver. $\tilde{E}_d(t)$ is the envelope of the driver only related with time
\begin{equation}
\tilde{E}_d(t)=\frac{\tilde{E}_d^{max}}{1+(\frac{t-t_0}{\frac{1}{2}t_0})^{10}},
\end{equation}
where $\tilde{E}_d^{max}=eE_d^{max}\lambda_{De}/T_e$ is the maximum amplitude of the driver, $t_0$ is the duration time of the peak driving electric filed. The form of the envelop of the driver can ensure that the waves are driven slowly by the driver up to a finite amplitude, thus a so-called \textquotedblleft adiabatic" distribution of resonance particles is formed.\cite{Dewar, (9)} Otherwise, if the wave is initialized suddenly with a large amplitude, a so-called \textquotedblleft sudden" distribution of resonance particles will be formed.\cite{Dewar,(9),Morales}

\begin{figure}[!tp]
	
	\includegraphics[width=1.0\columnwidth]{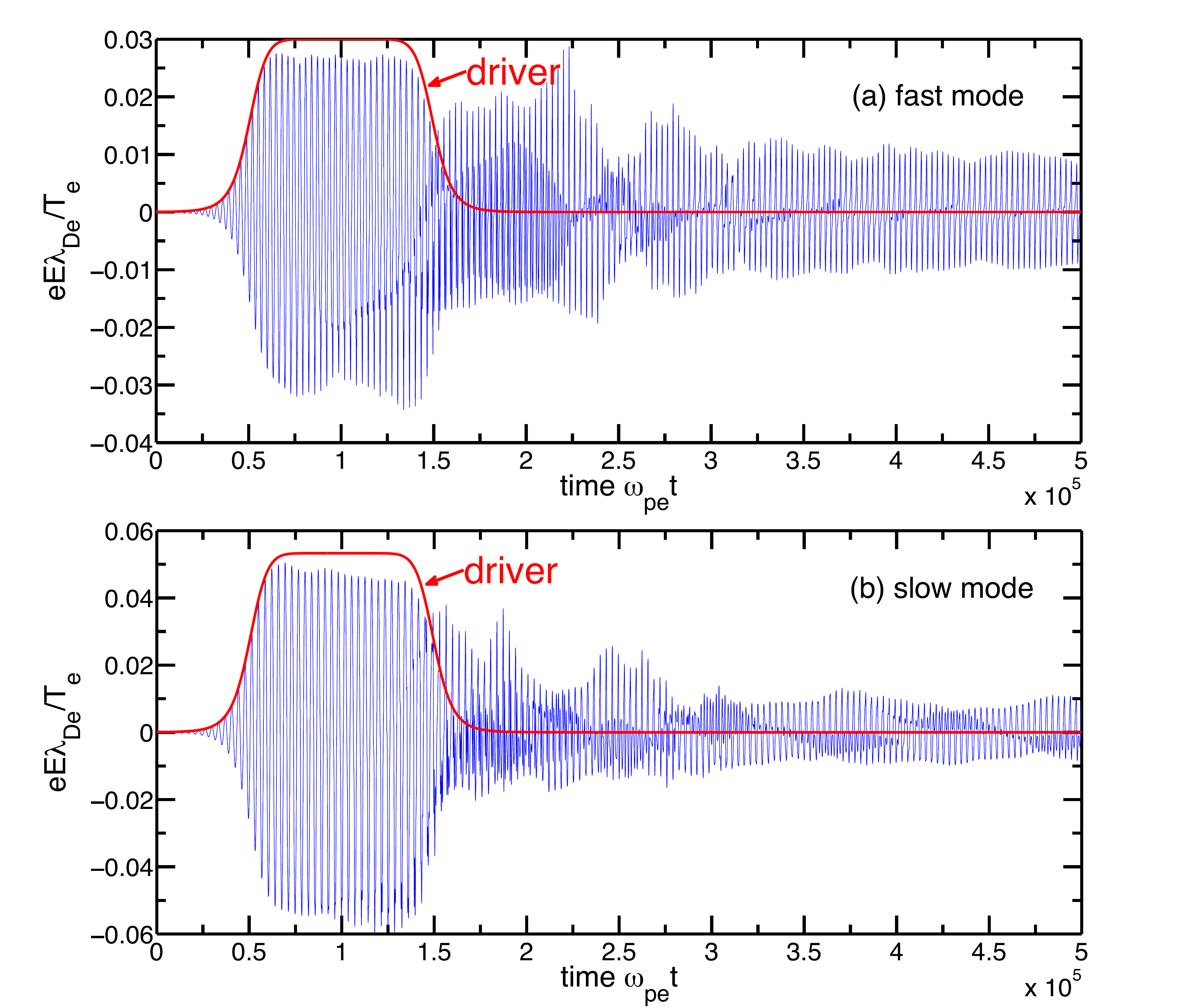}
	
	\caption	{\label{Fig:Excitation}(Color online) Time evolution of the electric field, calculated at a fixed point $x_0=5\lambda_{De}$, for (a) $T_i/T_e=0.1$, the fast IAW mode (the frequency of the driver $\omega_d=1.965\times10^{-3}\omega_{pe}$, the maximum amplitude of the driver $eE_d^{max}\lambda_{De}/T_e=0.03$), (b) $T_i/T_e=0.5$, the slow IAW mode ($\omega_d=1.75\times10^{-3}\omega_{pe},  eE_d^{max}\lambda_{De}/T_e=0.0533$) in the condition of $k\lambda_{De}=0.1$. The red line is the envelop of the driver.}
\end{figure}

\begin{figure}[!tp]
	\includegraphics[width=1.0\columnwidth]{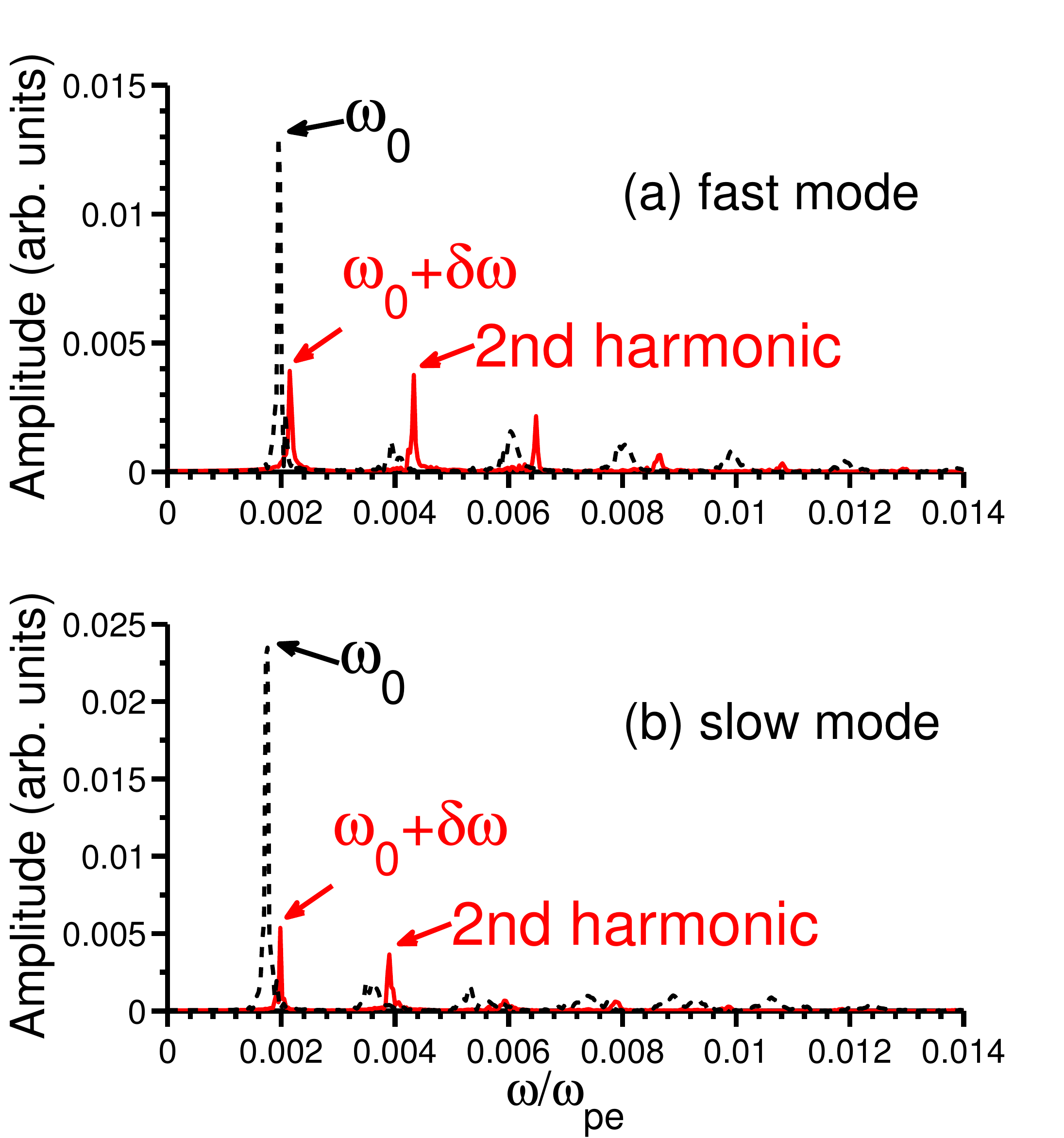}
	
	\caption{\label{Fig:Harmonic}(Color online) The frequency spectra of the electric field at $x_0=5\lambda_{De}$ and $\omega_{pe}t\in[0, 2\times10^{5}]$ (black dashed line), $\omega_{pe}t\in[3\times10^{5}, 4.9\times10^{5}] $(red solid line) for (a) the fast mode and (b) the slow mode. Corresponding to Fig. \ref{Fig:Excitation}. }
\end{figure}

In this paper, the fundamental frequency of the IAW calculated by the linear dispersion relation is chosen as the driver frequency, i.e., $\omega_d=\omega_L$ which is close to the resonant frequency of the small-amplitude nonlinear IAW when $k\lambda_{De}$ is small. The amplitude of the driver $\tilde{E}_d^{max}$ varies to excite different amplitude of nonlinear IAW. As shown in Fig. \ref{Fig:Excitation}, the examples of the driver amplitude $\tilde{E}_d^{max}=0.03$ for the fast mode and $\tilde{E}_d^{max}=0.0533$ for the slow mode are taken to excite the nonlinear IAW with the amplitude $\tilde{E}\approx0.01$. The external driving electric field is on from 0 to $2\times10^5\omega_{pe}^{-1}$ with the duration time $t_0=1\times10^{5}\omega_{pe}^{-1}$ which is much longer than the bounce time of the particles $\tau_{bj}=2\pi/\sqrt{q_jkE/m_j}$ so to ensure the excitation of the nonlinear IAW. After the driver is off, the electric field oscillates at almost constant amplitude, with a value that depends on the the driver amplitude $\tilde{E}_d^{max}$. We can notice that impulsive spikes appear in the electric field and the electric signal appears to be composed of many frequencies. Fourier analysis (shown in Fig. \ref{Fig:Harmonic}) will reveal the development of the harmonics when the driver is on and off, and also the frequency shift due to the harmonic generation.

\begin{figure}[!tp]
	\includegraphics[width=1.0\columnwidth]{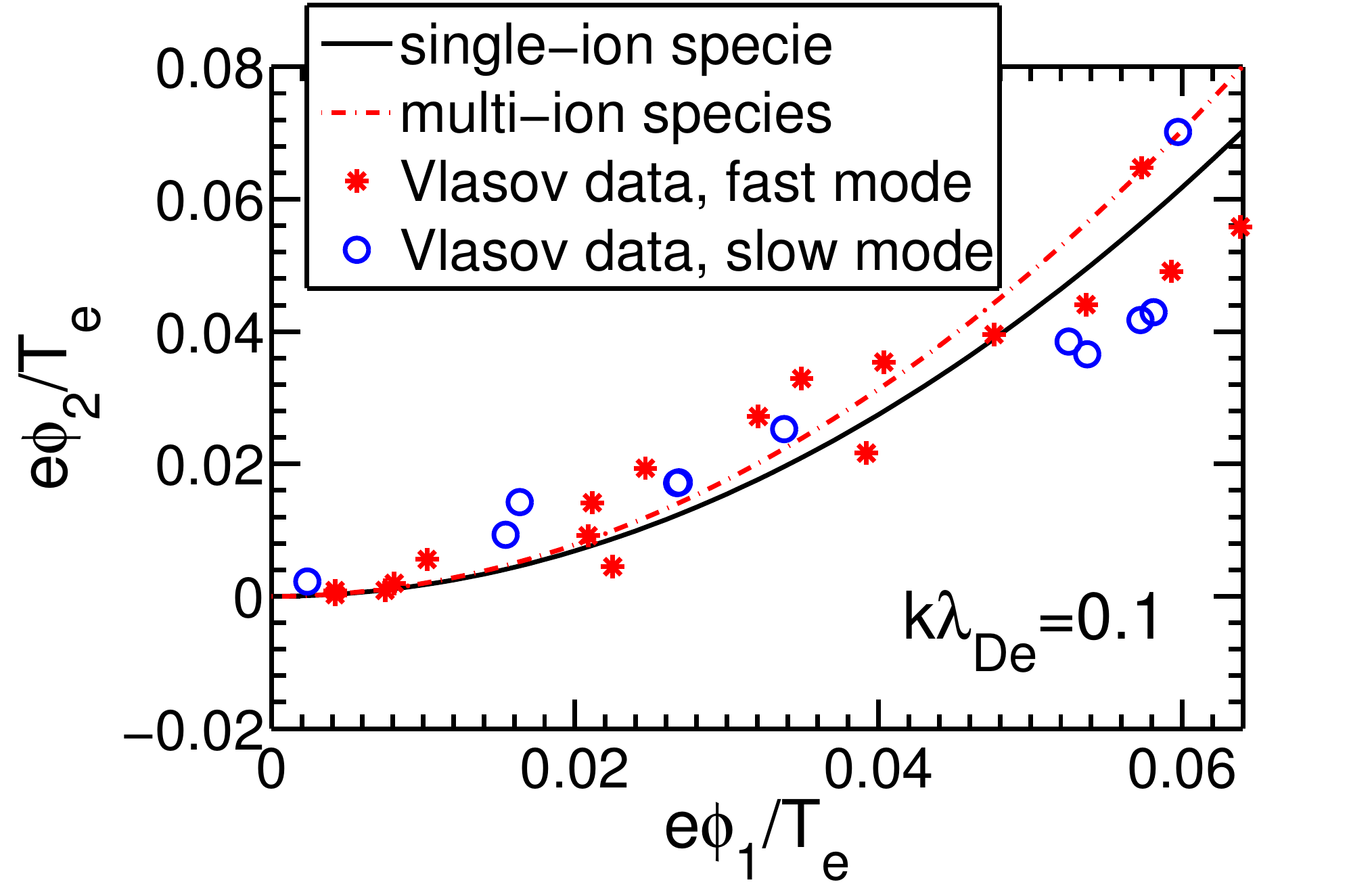}
	
	\caption{\label{Fig:phi_1-phi_2}(Color online) The relation of the second harmonic amplitude and the first harmonic amplitude for the fast mode and the slow mode. The simulation results are compared to the fluid theory including the cold ion fluid model of single-ion specie and multi-ion species.}
\end{figure}

\begin{figure*}[htd]
	\includegraphics[width=2\columnwidth]{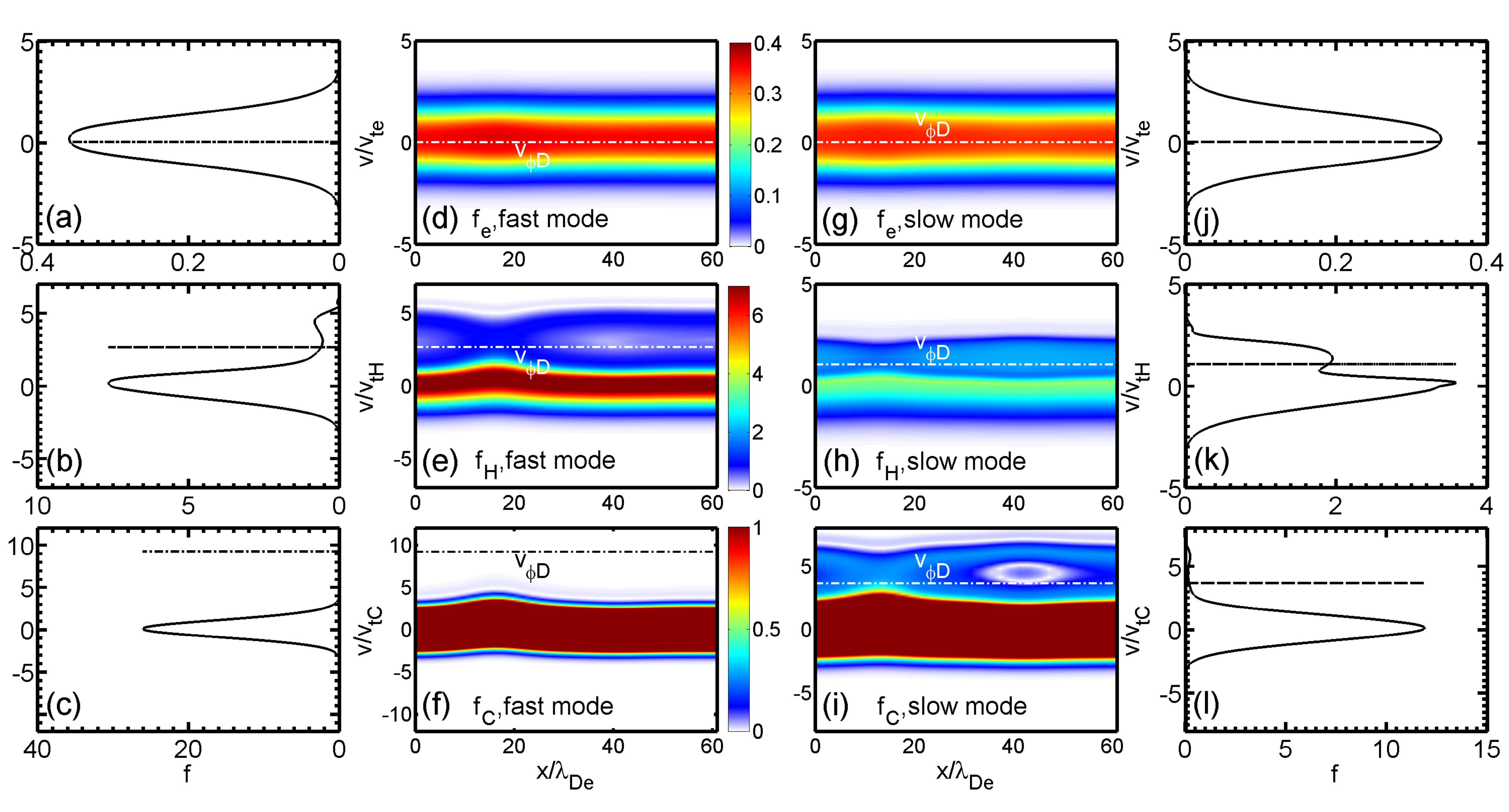}
	
	\caption{\label{Fig:PhasePicture}(Color online) The phase pictures and the corresponding distributions of electrons, H ions and C ions for the fast mode and the slow mode at the specific time $\omega_{pe}t_0=4\times10^5$, $\tilde{E}\approx0.01$. $v_{\phi D}$ is the phase velocity of the driver.}
\end{figure*}
\begin{figure}[htd]
	\includegraphics[width=\columnwidth]{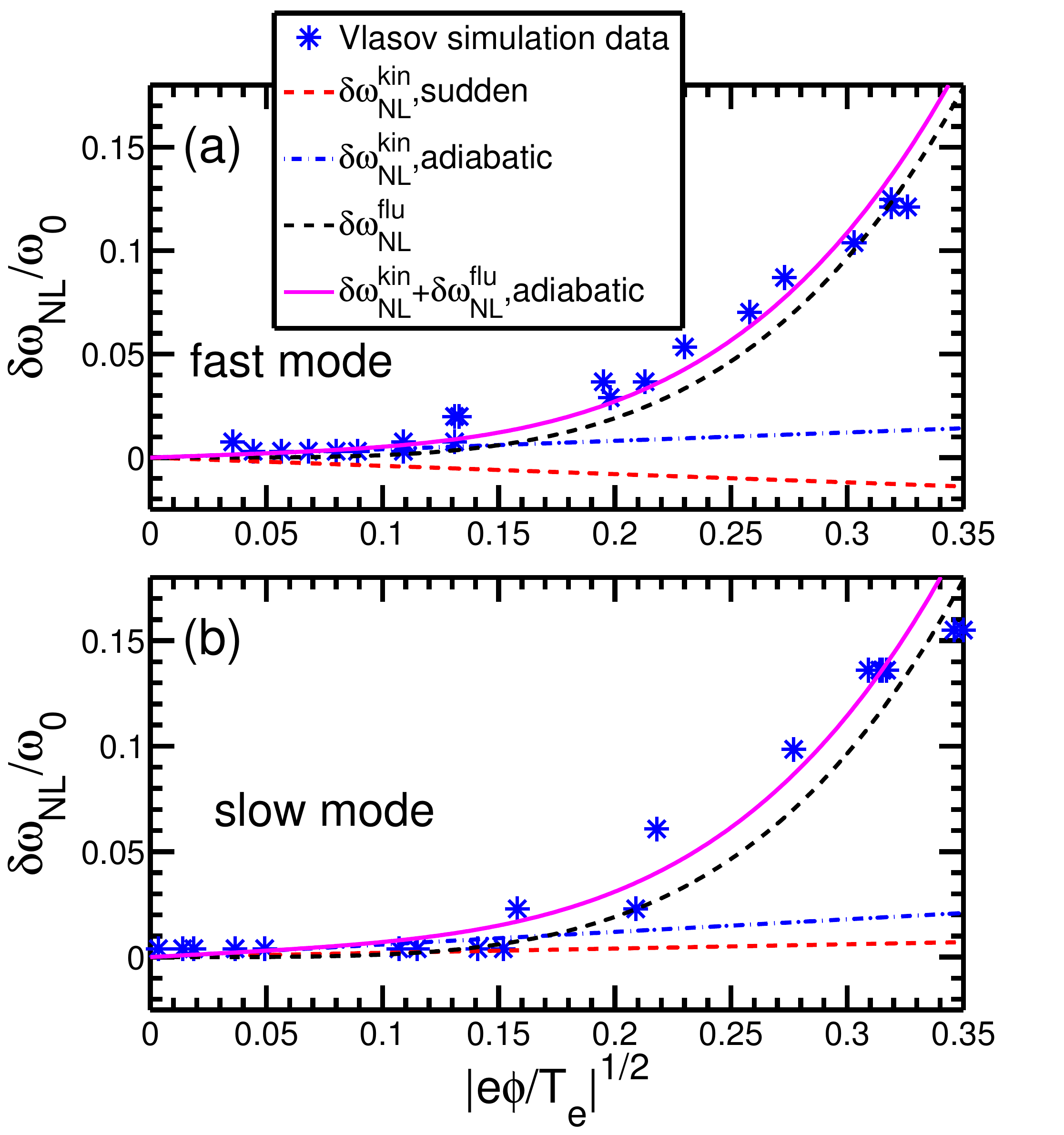}
	
	\caption{\label{Fig:NFS}(Color online) The nonlinear frequency shift (NFS) of (a) the fast mode, $T_i/T_e=0.1$ and (b) the slow mode, $T_i/T_e=0.5$ compared to fluid theory in the multi-ion species plasmas (Sec. \ref{Subsec:A. Fluid theory}) and the kinetic theory (Sec. \ref{Subsec:B. Kinetic theory}) in the condition of $k\lambda_{De}=0.1$, where \textquotedblleft sudden" and \textquotedblleft adiabatic" represent for the excitation condition of the ions (C or H ions), the excitation condition of the electrons is adiabatic, i.e., $\alpha_e=\alpha_{ad}$. }
\end{figure}

In Fig. \ref{Fig:Harmonic}, the frequency spectra analyses of the time period when the driver is on, i.e., $\omega_{pe}t\in[0, 2\times10^{5}]$, are shown as the black dashed lines which are labeled as $\omega_0$. When the driver is off and the electric field of IAW reaches constant amplitude, i.e., $\omega_{pe}t\in[3\times10^5, 4.9\times10^{5}]$, the frequency spectra are shown as the red solid lines which are labeled as $\omega_0+\delta\omega$. When the driver is on, the amplitude of the fundamental frequency $\omega_0$ is much larger than that of the harmonics. In this period, the harmonic is weak and the frequency $\omega_0$ is near the driver frequency $\omega_d$. However, when the driver is off and the electric field of the nonlinear IAW reaches constant finite amplitude, the harmonics especially the second harmonics grow and reach an amplitude level comparable to that of the fundamental mode driven by the driver. In this period, the frequency of the fundamental modes will shift with a quantity of $\delta\omega$ relative to $\omega_0$. The same process appears for the fast mode and the slow mode. These cases are taken to qualitatively show the nonlinear frequency shift of the IAW at the IAW electric field amplitude $\tilde{E}\approx0.01$ corresponding to the condition of Fig. \ref{Fig:Excitation}. The quantitatively calculations will be shown in Fig. \ref{Fig:NFS}. While the quantity of the frequency shift $\delta\omega$ is related to not only the amplitude of the nonlinear IAW but also the wave number $k\lambda_{De}$, which will be discussed in Sec. \ref{Sec:Discussions}. 

Fig. \ref{Fig:phi_1-phi_2} shows that the 2nd harmonic amplitude $e\phi_2/T_e$ varies with the fundamental mode amplitude $|\tilde{\phi_1}|$ (or the first harmonic amplitude, $\tilde{\phi_1}=e\phi_1/T_e$, corresponding to the mode labeled as $\omega_0+\delta\omega$ in Fig. \ref{Fig:Harmonic}). With the fundamental mode amplitude $|\tilde{\phi_1}|$ increasing, the harmonics especially the 2nd harmonic will increase correspondingly, which is consistent to the theoretical result of Fig. \ref{Fig:single-multi}(b) from Eq. \ref{Eq:phi_2-phi_1} in the appropriate range of $\tilde{\phi_1}$. However, if the fundamental mode amplitude $\tilde{\phi_1}$ is too large ($\tilde{\phi_1}\gtrsim0.06)$, the theoretical curve from Eq. \ref{Eq:phi_2-phi_1} will fail to fit the Vlasov simulation datas well due to the large amplitude higher-order harmonics such as the third even the fourth harmonics generation. Thus the multi-ion species fluid NFS model should be applied in the appropriate range of IAW amplitude. In the condition of the small wave number, such as $k\lambda_{De}=0.1$, the harmonic effect is obvious especially when the nonlinear IAW amplitude is large. As a result, the fluid NFS from harmonic generation will dominate in the total NFS. In our paper and also in Chapman's research\cite{T. Chapman_PRL}, the multi-ion species plasmas are considered. Therefore, the multi-ion species fluid NFS should be taken rather than the single-ion specie fluid NFS that was taken in Chapman's research. The detail discussions are given as shown in Sec. \ref{Sec:Discussions}.

The total nonlinear frequency shift comes from the harmonic generation (fluid) and the particles trapping (kinetic). As shown in Fig. \ref{Fig:PhasePicture}, the distributions and the phase pictures of particles for the fast mode and the slow mode are given to clarify the wave and particles interaction, thus the nonlinear frequency shift from particles trapping. After a certain number of bounce times, a quasi-steady, BGK-like state\cite{BGK} is thus established. For the phase velocities of both the fast mode and the slow mode are lower than the thermal velocity of electrons, thus the wave will gain energy from the electrons, as a result, electrons will provide a positive frequency shift of the nonlinear IAW. The contribution of H ions and C ions to the NFS of the fast mode and the slow mode is different for the different phase velocity of the modes compared to the ions thermal velocity. For the fast mode, the phase velocity is 2-3 times of the thermal velocity of H ions and much larger than that of C ions, thus the wave will give energy to H ions by trapping and can nearly not trap C ions, as a result, H ions will provide a negative NFS while C ions make no contribution to the NFS of the fast mode. However, for the slow mode, the phase velocity is close to the thermal velocity of H ions and 3-4 times of that of C ions, the wave and H ions will nearly not exchange energy while the wave will give energy to C ions by trapping, thus H ions nearly make no contribution to the NFS while C ions provide a negative NFS of the slow mode. This is the explanation of kinetic nonlinear frequency shift (KNFS) from each species for both the fast mode and the slow mode in the view of energy exchange of the wave and particles, which is consistent to the explanation of KNFS by Chapman\cite{T. Chapman_PRL} through theoretical analysis.

In Fig. \ref{Fig:NFS}, the Vlasov simulation datas of the NFS of nonlinear IAWs in the condition of $k\lambda_{De}=0.1$ is given when the electric field amplitude of nonlinear IAW varies. The fundamental frequency of the nonlinear IAW $\omega_0$ in the Vlasov datas is chosen as the frequency calculated by the dispersion relation with no damping $Re(\epsilon_L(Re(\omega), k))=0$. In the condition of $k\lambda_{De}=0.1$, the fluid NFS from harmonic generation is larger than the kinetic NFS from particles trapping when $|e\phi/T_e|^{1/2}\gtrsim0.15$. And the total NFS, $\delta\omega_{NL}/\omega_0$, can reach as large as nearly $15\%$ when the nonlinear IAW amplitude $|e\phi/T_e|$ reaches 0.1 in which the Vlasov simulation datas match the theory analytical curve ($\delta\omega_{NL}^{kin}+\delta\omega_{NL}^{flu}$) well. These results indicate the total NFS can reach a relative large value when the wave number $k\lambda_{De}$ is small for the reason that the NFS from harmonic generation dominants and increases quickly with the electric field amplitude, which is much larger than that researched by T. Chapman\cite{T. Chapman_PRL} in the condition of $k\lambda_{De}=1/3$. The comparison of the total NFS of the modes in the condition of $k\lambda_{De}=0.1$ and $k\lambda_{De}=1/3$ will be shown in Fig. \ref{Fig:NFS_k}. The Vlasov simulation datas are consistent to the curve of the total NFS including fluid NFS and kinetic NFS, $\delta\omega_{NL}^{flu}+\delta\omega_{NL}^{kin}$, where adiabatic is taken as the excitation condition of the ions and the fluid NFS analysis $\delta\omega_{NL}^{flu}$ is given by the multi-species cold ions fluid model. This verifies that the theoretical model of the fluid NFS in multi-ion species plasmas is valid in appropriate scale of $\tilde{\phi}$ (not too large). In the research of T. Chapman, the theory analysis of NFS from harmonic generation used the single-specie cold ion fluid model which was derived by Berger\cite{(9)}. It was clear that calculations of $\delta\omega_{NL}^{kin}$ from the kinetic NFS theory matched Vlasov datas well for low $\tilde{\phi}$, in which the kinetic NFS dominated, however, underestimated the total NFS $\delta\omega_{NL}$ at higher amplitudes of $\tilde{\phi}$ for the reason that harmonic generation in multi-ion species plasmas will contribute a further positive frequency shift than single-ion specie plasmas as shown in Fig. \ref{Fig:NFS_Chapman}. The comparison of the fluid NFS (or the total NFS) in the multi-ion species plasmas and the single-ion specie plasmas in the condition of Ref. \cite{T. Chapman_PRL} will be shown in Fig. \ref{Fig:NFS_Chapman}.

However, when the nonlinear IAW amplitude is too large, the higher-order harmonics such as the third even the fourth harmonic will be excited resonantly to a large amplitude, thus the theoretical model of the fluid NFS from harmonic generation in multi-ion species plasmas should retain the higher order of the variables in a Fourier series.

\section{\label{Sec:Discussions}Discussions}
\begin{figure}[htd]
	\includegraphics[width=\columnwidth]{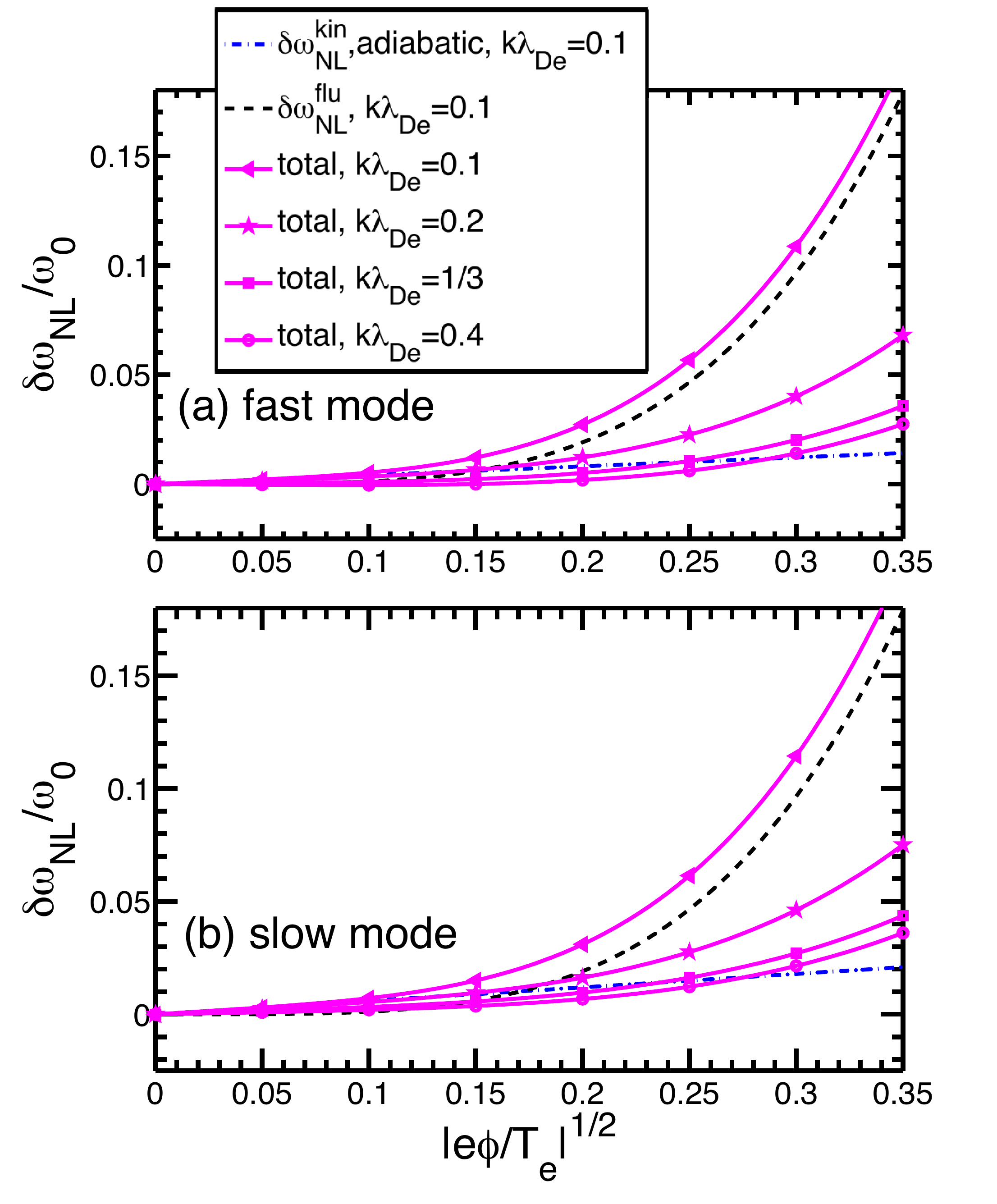}
	
	\caption{\label{Fig:NFS_k}(Color online) The analytical results of the total nonlinear frequency shift of (a) the fast mode, $T_i/T_e=0.1$ and (b) the slow mode, $T_i/T_e=0.5$ for various values of $k\lambda_{De}$, where \textquotedblleft adiabatic\textquotedblright   is considered as the excitation condition of the ions (C or H ions). \textquotedblleft total" presents for the total NFS, i.e., the kinetic NFS plus the multi-ion species fluid NFS $\delta\omega_{NL}^{kin}+\delta\omega_{NL}^{flu}, multi-ion\ species$. }
\end{figure}

Fig. \ref{Fig:NFS_k} shows the total NFS (fluid plus kinetic) of CH plasmas (C:H=1:1) as a function of wave amplitude for various values of the wave number $k\lambda_{De}$. We can see that the total NFS in the condition of $k\lambda_{De}=0.1$ is much larger than that in the condition of $k\lambda_{De}=1/3$ which is the condition in Ref. \cite{T. Chapman_PRL}. The reason is that for small $k\lambda_{De}$, harmonic generation plays a prominent role in NFS. With $k\lambda_{De}$ increasing, the fluid NFS decreases obviously as shown in Fig. \ref{Fig:single-multi}(c), thus the total NFS decreases correspondingly.

\begin{figure}[htd]
	\includegraphics[width=\columnwidth]{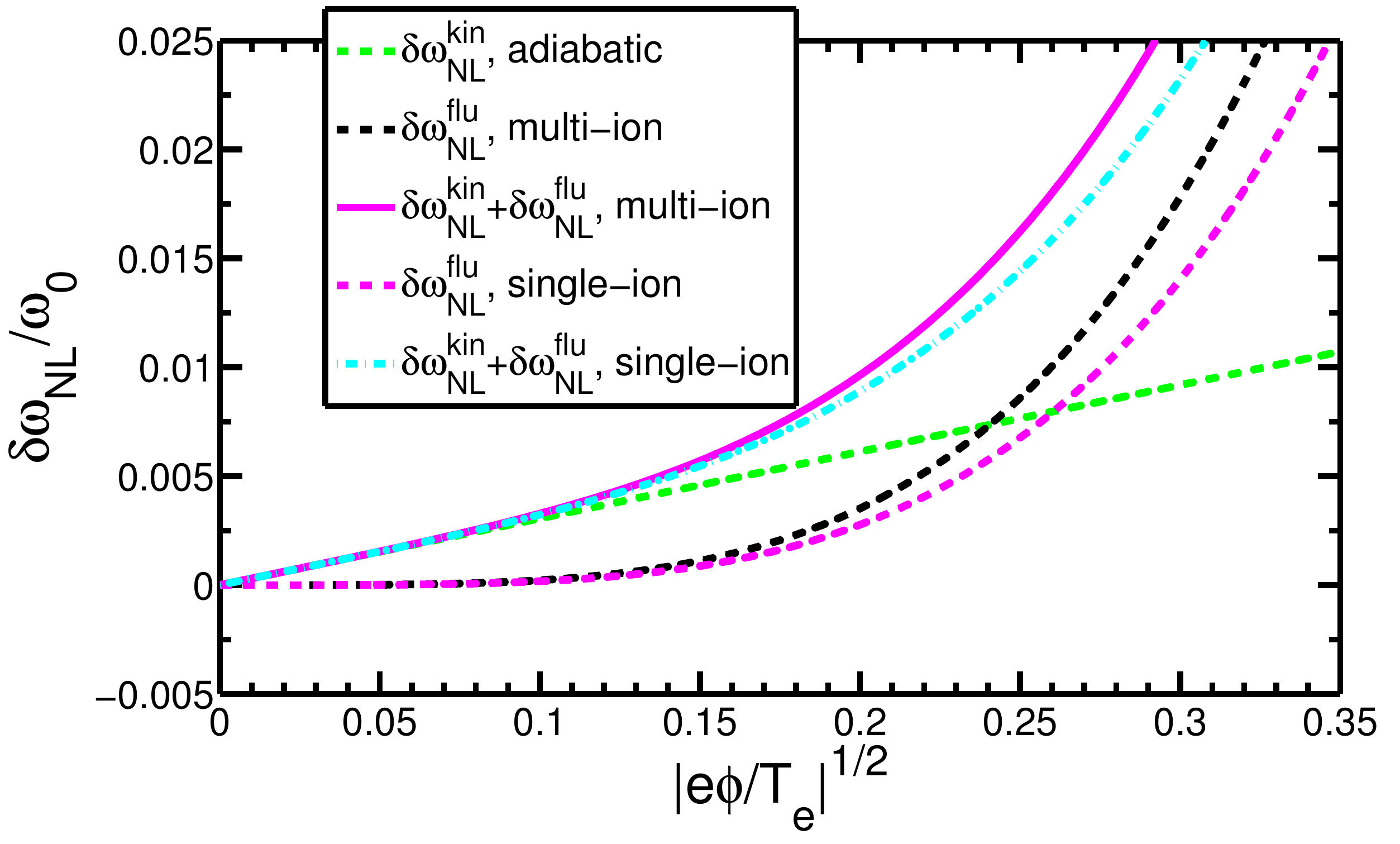}
	
	\caption{\label{Fig:NFS_Chapman}(Color online) The comparison of the multi-ion species fluid NFS (the total NFS: $\delta\omega_{NL}^{kin}+\delta\omega_{NL}^{flu}, multi-ion$) and the single-ion specie fluid NFS (the total NFS: $\delta\omega_{NL}^{kin}+\delta\omega_{NL}^{flu}, single-ion$) for the slow mode in the condition of Ref. \cite{T. Chapman_PRL}, $T_i/T_e=0.5, k\lambda_{De}=1/3$, corresponding to Fig. 3(d) in Ref. \cite{T. Chapman_PRL}.}
\end{figure}

To compare the multi-ion species fluid NFS and Chapman's research\cite{T. Chapman_PRL}, the condition of the slow mode $T_i/T_e=0.5, k\lambda_{De}=1/3$, which is the same as Ref. \cite{T. Chapman_PRL}, is taken as an example. The kinetic theory is the same as Chapman's research (as shown in Sec. \ref{Subsec:B. Kinetic theory}) and the excitation condition of the ions is \textquotedblleft adiabatic\textquotedblright. As shown in Fig. \ref{Fig:NFS_Chapman}, especially when the IAW amplitude is large, the multi-ion species fluid NFS (Sec. \ref{Subsec:A. Fluid theory}) is larger than the single-ion specie fluid NFS which was derived by Berger\cite{(9)} and made use of by Chapman in Ref. \cite{T. Chapman_PRL}. Thus, the total NFS (kinetic NFS plus multi-ion species fluid NFS) is larger than the total NFS in Ref. \cite{T. Chapman_PRL} (kinetic NFS plus single-ion specie fluid NFS) especially when the IAW amplitude is large. This result of multi-ion species fluid NFS will make the total NFS fit the Vlasov datas in Fig. 3(d) of Ref. \cite{T. Chapman_PRL} better especially at higher IAW amplitude. This fluid NFS model of multi-ion species plasmas is a correction to the research of Chapman et al. \cite{T. Chapman_PRL} for their research just considered the single-ion specie fluid NFS model that was derived by Berger\cite{(9)}. 

When $f_C=0, f_H=1$, i.e., in the H plasmas, the fluid NFS from the multi species cold ions fluid model is consistent to that from the single specie cold ion fluid model derived by Berger\cite{(9)}. This indicates that the multi-ion species NFS model is valid and can also cover the single-ion specie NFS model if we take $f_C=0$ or $f_H=0$. We can find the single-ion specie fluid NFS model is independent of the ion specie. However, the multi-ion species fluid NFS model is related to the ion specie and the number fraction of the ion $f_i$. This paper takes $f_C=f_H=0.5$ as an example, the fluid NFS of multi-ion species plasmas is larger than that of single-ion specie plasmas for the effect of mass and charge in multi-ion species plasmas, thus harmonic generation in multi-ion species plasmas contributes a further positive frequency shift. This multi-ion species fluid NFS model gives the explanation of why the calculations of total NFS in Fig. 3(d) of Ref. \cite{T. Chapman_PRL} underestimate the real NFS $\delta\omega_{NL}$ at higher IAW amplitudes.

\section{\label{Sec:Conclusions}Conclusions}
In summary, the nonlinear frequency shift from harmonic generation derived from the multi-species cold ions fluid model has been given to calculate the fluid NFS for the first time and verified to be consistent to the Vlasov results for both the fast mode and the slow mode. 
And this multi-ion species fluid NFS model can be applied to many multi-ion species plasmas. 
The pictures of the NFS of nonlinear IAW from the harmonic generation and the particles trapping are shown to explain the NFS process. The fluid NFS from harmonic generation is related to not only the nonlinear IAW amplitude but also the wave number $k\lambda_{De}$.
When the wave number $k\lambda_{De}=0.1$, the fluid NFS from harmonic generation dominants in the scale of $|e\phi/T_e|^{1/2}\gtrsim0.15$ and will reach as large as $\sim15\%$ when $|e\phi/T_e|^{1/2}\sim0.35$ in which the theory analytical calculations are consistent to the Vlasov simulation results. This indicates that the fluid NFS dominants in the saturation of SBS in the condition of small $k\lambda_{De}$ especially when the wave amplitude is large.\\

\begin{acknowledgments}
We are pleased to acknowledge useful discussions with K. Q. Pan. This research was supported by the National Natural Science Foundation of China (Grant Nos. 11575035, 11475030 and 11435011) and National Basic Research
Program of China (Grant No. 2013CB834101).
\end{acknowledgments}


\end{document}